\documentclass[12pt,cls,onecolumn]{IEEEtran}
\usepackage{graphicx,amsmath,amssymb,epsfig, amsfonts, cite, latexsym, cuted, multicol, multirow, subfigure, stfloats, array, tabularx}
\usepackage{subeqnarray}
\usepackage{color}
\usepackage{setspace}
\usepackage{anysize}

\begin{document}

\title{Network-Decomposed Hierarchical Cooperation in Ad Hoc Networks With Social Relationships}
\author{Cheol~Jeong,~\IEEEmembership{Member,~IEEE,}
        and~Won-Yong~Shin,~\IEEEmembership{Senior Member,~IEEE}
\thanks{The work of W.-Y.~Shin was
supported by the Basic Science Research Program through the
National Research Foundation of Korea (NRF) funded by the Ministry
of Education (2017R1D1A1A09000835). The work of C. Jeong was supported by the NRF
through the Korea Government under Grant NRF-2017R1C1B1009145. {\em (Corresponding author:
Won-Yong Shin.)}}
\thanks{C. Jeong is with the School of Intelligent Mechatronics Engineering, Sejong University, Seoul 05006, Republic of Korea. E-mail: cheol.jeong@ieee.org.}
\thanks{W.-Y. Shin is with the Department of
Computer Science and Engineering, Dankook University,
Yongin 16890, Republic of Korea. E-mail: wyshin@dankook.ac.kr.}%
\thanks{Manuscript received September 24, 2017; revised June 16, 2018; accepted August 21, 2018.}
} \maketitle


\markboth{IEEE Transactions on Wireless Communications} {Jeong {et al.}:
Network-Decomposed Hierarchical Cooperation in Ad Hoc Networks With Social Relationships}


\newtheorem{definition}{Definition}
\newtheorem{theorem}{Theorem}
\newtheorem{lemma}{Lemma}
\newtheorem{example}{Example}
\newtheorem{corollary}{Corollary}
\newtheorem{proposition}{Proposition}
\newtheorem{conjecture}{Conjecture}
\newtheorem{remark}{Remark}

\newcommand{\red}[1]{{\textcolor[rgb]{1,0,0}{#1}}}
\newcommand{\blue}[1]{{\textcolor[rgb]{0,0,1}{#1}}}
\newcommand{\Vgreen}[1]{{\textcolor[rgb]{0,0.5,0}{#1}}}

\def \diag{\operatornamewithlimits{diag}}
\def \log{\operatorname{log}}
\def \rank{\operatorname{rank}}
\def \out{\operatorname{out}}
\def \exp{\operatorname{exp}}
\def \arg{\operatorname{arg}}
\def \E{\operatorname{E}}
\def \tr{\operatorname{tr}}
\def \SNR{\operatorname{SNR}}
\def \dB{\operatorname{dB}}
\def \ln{\operatorname{ln}}

\def \be {\begin{eqnarray}}
\def \ee {\end{eqnarray}}
\def \ben {\begin{eqnarray*}}
\def \een {\end{eqnarray*}}

\newcommand{\Pro}[1]{\mathrm{Pr}\left\{#1\right\}}
\newcommand{\LIF}[2]{\tilde{L}_{\pi_1(#1),#2}}
\newcommand{\TIL}[2]{L_{\pi_2(#1),#2}}
\newcommand{\TIF}[2]{T_{\pi_1(#1),#2}}
\newcommand{\KIF}[2]{T_{\pi_1(#1),\pi_2(#2)}}
\newcommand{\snr}{\textsf{snr}}
\newcommand{\sinr}{\textsf{sinr}}
\newcommand{\CanSB}{\mathcal{B}}
\newcommand{\CanSA}{\mathcal{A}}
\newcommand{\Norm}[1]{\left|{#1}\right|}
\newcommand{\PL}{\textsf{PL}}

\begin{abstract}
In this paper, we introduce a {\em network-decomposed}
hierarchical cooperation (HC) protocol and completely characterize
the corresponding throughput--delay trade-off for a large wireless
ad hoc network formed in the context of {\em social
relationships}. Instead of randomly picking source--destination
pairings, we first consider a distance-based social formation
model characterized by the social group density $\gamma$ and the
number of social contacts per node, $q$, where the probability
that any two nodes in distance $d$ away from each other are
socially connected is assumed to be proportional to $d^{-\gamma}$,
which is a feasible scenario. Then, using muiltihop and
network-decomposed HC protocols under our social formation model,
we analyze a generalized throughput--delay trade-off according to
the operating regimes with respect to parameters $\gamma$ and $q$
in both a dense network of unit area and an extended network of
unit node density via a non-straightforward network transformation
strategy. Our main results reveal that as $\gamma$ increases,
performance on the throughput--delay trade-off can remarkably be
improved, compared to the network case with no social
relationships. It is also shown that in the dense network, the
network-decomposed HC protocol always outperforms the multihop
protocol, while the superiority of the network-decomposed HC
depends on $\gamma$ and the path-loss exponent in the extended
network.
\end{abstract}


\newpage

\section{Introduction}
Communications between users (i.e., a source and a destination)
over a wireless network usually take place based on {\it
friendship}, which is defined as online or offline social
relationships among users. In other words, a source and its
destination(s) are not just randomly paired in real-world
communications, and rather a source tends to select its
destination(s) along with friendships.
In~\cite{LataneLiuNowakBoneventoZheng:95,Kleinberg:Nature00,BackstromSunMarlow:WWW10,Liben-NowellNovakKumarRaghavanTomkins:05,ShinSinghChoEverett:JIS15},
it was observed that social interactions among users indeed depend
heavily on the {\it geographic proximity} of them.
In~\cite{Liben-NowellNovakKumarRaghavanTomkins:05}, a close
relationship between geographic distance and probability
distribution of friendship was demonstrated by experimental
results based on the LiveJournal social network. More
specifically, it was shown that the probability of befriending a
particular user is inversely proportional to the positive power of
the geographic
distance~\cite{Liben-NowellNovakKumarRaghavanTomkins:05}.
In~\cite{ShinSinghChoEverett:JIS15}, the degree of friendship
related to the issue of space was further studied on
Twitter---the number of friends according to
distance follows a double power-law distribution on Twitter,
indicating that the probability of befriending a particular
Twitter user is significantly reduced beyond a certain geographic
distance between users. Moreover, there have been extensive
studies on understanding the nature of friendships with respect to
the geographic distance in large-scale online social networks such
as Twitter~\cite{KwakLeeParkMoon:10},
Facebook~\cite{ViswanathMisloveChaGummadi:09},
Flickr~\cite{MisloveKoppulaGummadiDruschelBhattacharjee:08},
LiveJournal~\cite{Liben-NowellNovakKumarRaghavanTomkins:05}, and
Foursquare~\cite{ChenZhuangCaoHui:14}, while validating the
small-world phenomenon and scale-free degree distributions. On the
other hand, it has widely been known that social relationships
influence users' interactions with each other in physical space;
thus, users' social ties are closely related to the interactions
of users' communication devices subject to diverse physical
coupling (see, e.g.,~\cite{ChenChiangPoor:JSAC13,Gongetal:ToN17}
and references therein). For example, users at
close proximity in a social group can share their pictures and
videos, play games with friends, or exchange files using
device-to-device (D2D)
communication~\cite{DatsikiAntonopoulosZorbaVerikoukis:Access16}.
The D2D communications between firefighters or between police
officers for public safety are another examples of wireless social
networks in which users are socially tied and their geographic
distances are short. Another application of wireless social
networks includes content-centric (caching)
communications~\cite{GitzenisPaschosTassiulas:13,MalikLimShin:18}
that content objects are cached by numerous nodes over a network,
in which each request is served by nearby content source nodes in
a friendship relation. For this reason, there has been a growing
interest in analyzing the impact of social groups on the
performance of wireless networks.
In~\cite{ZhangPanSongSaadDawyHan:TWC15,ZhaoLiCaoJiangGe:TWC15,DatsikiAntonopoulosZorbaVerikoukis:Access16,LiWuHuiJinChen:14},
traffic offloading, resource allocation, and medium access control
(MAC) protocols were designed for device-to-device wireless
communications in a social-aware perspective. In~\cite{LiHuiJinSuZeng:10}, the impact of social
selfishness on the performance of epidemic routing was also
investigated in delay tolerant networks. Social context-aware
small cell networks were designed
in~\cite{SemiariSaadValentinBennisPoor:15} by optimizing the
overall allocation of resources. In~\cite{AlimPanThaiSaad:17}, it
was presented how to form multihop D2D connections based on a
community-based approach in D2D communications. Moreover, the
throughput scaling laws of large wireless {\it ad hoc} networks
were also studied by incorporating the notion of social
characteristics into their network
models~\cite{AzimdoostSadjadpourGarcia-Luna-Aceves:TWC13,KiskaniAzimdoostSadjadpour:TWC16,WangShaoLiYangLiJiang:TPDS15,HouChengLiShengLui:TNET},
where the throughput scaling results depend on the number of nodes
and the geographic distance unlike the case with no social
relationships. In this paper, we aim to characterize a fundamental
throughput--delay trade-off of a large wireless ad hoc network,
where users communicate with others in the context of {\em social
relationships}.

\subsection{Related Work}

In~\cite{GuptaKumar:00}, it was shown that the aggregate
throughput of a large wireless network having $n$
source-destination (S--D) pairs randomly distributed in a unit
area (i.e., a dense network) scales as $\Theta(\sqrt{n/\log n})$,
which is achieved by the nearest-neighbor multihop (MH) protocol.
This throughput scaling was improved to $\Theta(\sqrt{n})$ using
percolation theory~\cite{FranceschettiDouseTseThiran:07}. MH
protocols were further studied and analyzed in various
aspects~\cite{GuptaKumar:03,XueXieKumar:05,ShinChungLee:TIT13}.
There has been a great deal of research to improve the aggregate
throughput of dense networks up to a linear scaling
in~\cite{OzgurLevequeTse:07,NiesenGuptaShah:09,ZemlianovVeciana:05,O.Dousse:INFOCOM02,ShinJeonDevroyeVuChungLeeTarokh:08,GrossglauserTse:02,ZhangXuWangGuizani:TC10,LiZhangFang:TMC11,YoonShinJeon:TMC17}.
It was shown that an almost linear throughput scaling, i.e.,
$\Theta(n^{1-\epsilon})$ for an arbitrarily small $\epsilon>0$,
can be achieved by hierarchical cooperation (HC)
protocols~\cite{OzgurLevequeTse:07,NiesenGuptaShah:09}. The impact
and benefits of infrastructure support in improving the throughput
scaling in hybrid networks, consisting of both wireless ad hoc
nodes and infrastructure nodes, were studied
in~\cite{ZemlianovVeciana:05,O.Dousse:INFOCOM02,ShinJeonDevroyeVuChungLeeTarokh:08}.
Novel techniques such as networks with node
mobility~\cite{GrossglauserTse:02} and directional
antennas~\cite{ZhangXuWangGuizani:TC10,LiZhangFang:TMC11,YoonShinJeon:TMC17}
were also introduced to achieve a linear throughput scaling.
Besides the throughput, delay is also a key performance metric in
most wireless network applications. One can usually improve the
per-node throughput at the cost of an increased delay of a packet.
The trade-off between throughput and delay metrics of both static
and mobile ad hoc networks was examined in terms of scaling laws
in some
papers~\cite{NeelyModiano:05,ElGamalMammenPrabhakarShah:TIT06,ElGamalMammenPrabhakarShah:TIT06-2,OzgurLeveque:TIT10}.
In~\cite{NeelyModiano:05}, the throughput--delay trade-off of a
mobile ad hoc network adopting a two-hop relay protocol was
analyzed under a simple independent and identically distributed
(i.i.d.) mobility model.
In~\cite{ElGamalMammenPrabhakarShah:TIT06,ElGamalMammenPrabhakarShah:TIT06-2},
the throughput--delay trade-off was derived in another mobile
network adopting a random walk mobility model as well as in a
static network. In~\cite{OzgurLeveque:TIT10}, the
throughput--delay trade-off of a static ad hoc network was studied
by modifying the original HC protocol in~\cite{OzgurLevequeTse:07}
in order to improve the delay performance for the same throughput.

In all these previous studies, it was assumed that a source
selects its destination {\em at random} for analytical
convenience. In practice, however, this assumption is hardly
realistic since a source and its destination tend to be paired up
along with one-to-one friendship in the presence of social groups.
Thus, existing achievable schemes and analytical frameworks that
show the capacity scaling laws cannot be directly applicable to
the performance analysis of ad hoc networks with social groups.
Recently, the notion of social relationships was taken into
account in studying the capacity scaling laws of large wireless ad
hoc
networks~\cite{AzimdoostSadjadpourGarcia-Luna-Aceves:TWC13,KiskaniAzimdoostSadjadpour:TWC16,WangShaoLiYangLiJiang:TPDS15,HouChengLiShengLui:TNET}.
In~\cite{AzimdoostSadjadpourGarcia-Luna-Aceves:TWC13}, the
throughput scaling achieved by the MH protocol was analyzed again
under a social formation model such that each node has a social
group consisting of a fixed number of nodes and selects its
destination uniformly among the nodes in its social group. This
result was generalized in~\cite{KiskaniAzimdoostSadjadpour:TWC16}
by assuming social groups with different numbers of nodes and a
non-uniform probability of selecting one destination in each
social group. As an alternative approach to analyzing the network
throughput scaling, it was assumed
in~\cite{WangShaoLiYangLiJiang:TPDS15} that the number of friends
(i.e., the friendship degree) follows a Zipf's
distribution~\cite{ManningSchutze:99} and the probability of
befriending a particular user depends on both the geographic
distance between nodes and the node density.
In~\cite{HouChengLiShengLui:TNET}, the capacity scaling of a
hybrid network with social contact behavior was also investigated.

\subsection{Main Contributions}
In this paper, we introduce a new HC protocol, termed a {\em
network-decomposed} HC protocol, and characterize a general
throughput--delay trade-off in a large wireless ad hoc network
formed in the context of {\it social relationships}. To this end,
we first consider a distance-based social formation model
parameterized by the social group density $\gamma$, the number of
social contacts per node, $q$, and the probability that a source
selects its destination among social contacts. To better
illustrate our results, we identify three operating regimes
partitioned according to the social group density $\gamma$. More
specifically, we focus on designing the network-decomposed HC
protocol so that the network under our social formation model
operates properly, since the conventional HC protocols
in~\cite{OzgurLevequeTse:07,OzgurLeveque:TIT10} do not guarantee
the best trade-off performance under our network model. In the
proposed protocol, the whole network is divided into multiple
non-overlapping subnetworks, each of which operates in parallel
along with the HC protocol. A time-division multiple access (TDMA)
operation is used to avoid an edge node problem that may occur
with network decomposition. To be specific, when the network is
decomposed into multiple subnetworks, there may exist some sources
in a subnetwork, whose destinations are out of the subnetwork. To
solve this problem, we apply the 4-TDMA strategy with which all
the subnetworks are shifted to the right, shifted up, and shifted
diagonally in consecutive three time slots and remain unshifted in
one time slot. Using both the MH protocol
in~\cite{AzimdoostSadjadpourGarcia-Luna-Aceves:TWC13}, revisited
in the context of social relationships, and the network-decomposed
HC protocol and then computing the average distance between a
source and its destination derived under the social formation
model, we completely characterize the throughput--delay trade-off
according to each operating regime with respect to parameters
$\gamma$ and $q$ in not only a dense network of unit area but also
an extended network of unit node density through a
non-straightforward network transformation strategy in terms of
$\gamma$. As main results, we show that the throughput--delay
trade-off is significantly improved as $\gamma$ increases beyond a
certain value, compared to the network case with no social
relationships. We also show that in the dense network, the network-decomposed HC
protocol always outperforms the multihop protocol, while the
superiority of the network-decomposed HC is determined according
to $\gamma$ and the path-loss exponent in the extended network.

Our main contributions are five-fold and summarized as follows: \begin{itemize}
    \item We incorporate the notion of social relationships into a large wireless ad hoc network in order to analyze a fundamental throughput--delay trade-off for not only dense networks but also more challenging extended networks, whereas the throughput performance was only characterized in some prior studies dealing with wireless social networks~\cite{AzimdoostSadjadpourGarcia-Luna-Aceves:TWC13,KiskaniAzimdoostSadjadpour:TWC16,WangShaoLiYangLiJiang:TPDS15,HouChengLiShengLui:TNET}.
  \item Unlike the conventional HC protocols~\cite{OzgurLevequeTse:07,{OzgurLeveque:TIT10}} that did not exploit one-to-one social relationships, we propose a network-decomposed HC protocol that is properly designed based on the 4-TDMA strategy under the distance-based social formation model.
  \item To better interpret our results along with achievable schemes, we also identify three operating regimes on the throughput--delay trade-off with respect to the social group density and the number of social contacts.
  \item In both dense and extended networks with our social formation model, we completely characterize a general throughput--delay trade-off achieved by the MH and network-decomposed HC protocols according to each operating regime.
  \item Furthermore, we conduct numerical evaluation to validate that our analytical results show trends consistent with computer simulation results.
\end{itemize}

\subsection{Organization and Notations}
The rest of the paper is organized as follows. The system model
and preliminaries are given in Section~\ref{SEC:SystemModel}. The
main results are presented in Section~\ref{SEC:MainResults}. In
Section~\ref{SEC:TD-Tradeoff-HC}, the network-decomposed HC
protocol is described and the corresponding throughput--delay
trade-off is derived. In Section~\ref{SEC:Simulation}, numerical results are presented. Section~\ref{SEC:Conclusion} summarizes the
paper with some concluding remarks.

Throughout this paper,
$\mathbb{E}[\cdot]$ indicates the statistical expectation; and
$\Pr\{\mathcal{A}\}$ is the probability of an event $\mathcal{A}$.
We use the following asymptotic notation: i) $f(x)=O(g(x))$ means
that there exist positive constants $C$ and $c$ such that
$f(x)\leq Cg(x)$ for all $x>c$, ii) $f(x)=\Omega(g(x))$ if
$g(x)=O(f(x))$, iii) $f(x)=\omega(g(x))$ means that
$\lim_{x\rightarrow \infty}\frac{g(x)}{f(x)}=0$, and iv)
$f(x)=\Theta(g(x))$ if $f(x)=O(g(x))$ and $g(x)=O(f(x))$.

\section{Models and Problem Definition}\label{SEC:SystemModel}
In this section, we first describe not only our network model but
also our social formation model. We then present important lemmas
that are required to derive our main results including the
throughput--delay trade-off.

\subsection{System and Channel Models}

We consider the following two types of network configurations: 1)
a dense network of unit
area~\cite{GuptaKumar:00,ElGamalMammenPrabhakarShah:TIT06,ElGamalMammenPrabhakarShah:TIT06-2,OzgurLevequeTse:07}
and 2) an extended network of unit node
density~\cite{FranceschettiDouseTseThiran:07,OzgurLevequeTse:07}.
In what follows, the dense network model is assumed, and our main
results will be extended to the extended network configuration
later (refer to Sections~\ref{SEC:TD_HC}
and~\ref{SEC:ExtendedNetwork}). It is assumed that our network is composed of $n$ nodes, which are uniformly and independently distributed in a square of given unit area, and one central processor, which enables the network to be globally synchronized and coordinated. Specifically, the central processor plays a role of not only controlling interference based on the TDMA strategy but also determining packet delivery routes using geo-located information of nodes. Updated information is delivered to the central processor from nodes only when new S--D pairings are established. Each node acts as a source and has exactly one corresponding
destination node. Now, let us turn to the channel modeling. Under wireless networks in light-of-sight (LOS) environments, the complex channel
between nodes $i$ and $k$ is denoted by $h_{ki}=\frac{e^{j\theta_{k,i}}}{d_{k,i}^{\alpha/2}}$ where $\theta_{k,i}$ is the random phase uniformly distributed in
$[0,2\pi)$,\footnote{In~\cite{FranceschettiMiglioreMinero:09}, it
was shown that the capacity scaling of wireless networks is
fundamentally limited by the relation between $n$ and the
wavelength due to laws of physics. Based on this investigation,
the throughput scaling achieved by HC was rediscovered
in~\cite{LeeChung:12}. In this paper, however, we adopt the i.i.d.
phase assumption as
in~\cite{OzgurLevequeTse:07,NiesenGuptaShah:09,ShinJeonDevroyeVuChungLeeTarokh:08}
for analytical tractability. Moreover, we deal
simply with the LOS channel rather than multipath fading channels
to focus on analyzing the effect of social relationships in our
network. This is due to the fact that throughput scaling laws do
not fundamentally change in the presence of fading if all nodes
have their own traffic
demands~\cite{XueXieKumar:05,ShinChungLee:TIT13,NebatCruzBhardwaj:09,JovicicViswanathKulkarni:04}.}
$d_{k,i}$ denotes the distance between nodes $i$ and $k$, and
$\alpha\geq 2$ is the path-loss exponent. The phase $\theta_{k,i}$
and the path-loss model are based on a far-field assumption, i.e.
the distance between any two nodes is assumed to be much larger
than the carrier wavelength. It is also assumed
that one-to-one friendship relations on online social networks are
available at the central processor so that sources and
destinations are paired up in the presence of social groups.
Moreover, the central processor updates the social relationships
between any two nodes, which takes place intermittently since the
period of such changes to friendship would be much longer than the
entire communication period.

\subsection{Social Formation Model} \label{SEC:SocialModel}

\begin{figure}[t!]
  \centering
  \subfigure[$\gamma$ is small.]{
  \includegraphics[width=1.4in]{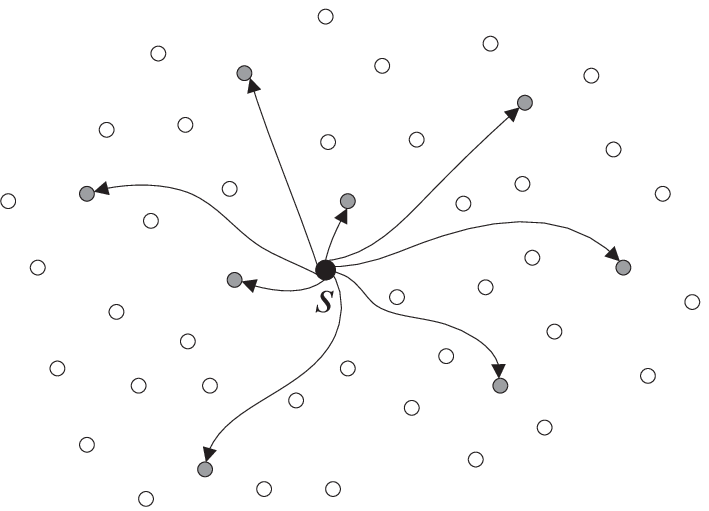}}
  \hspace{0.4cm}
  \centering
  \subfigure[$\gamma$ is large.]{
  \includegraphics[width=1.4in]{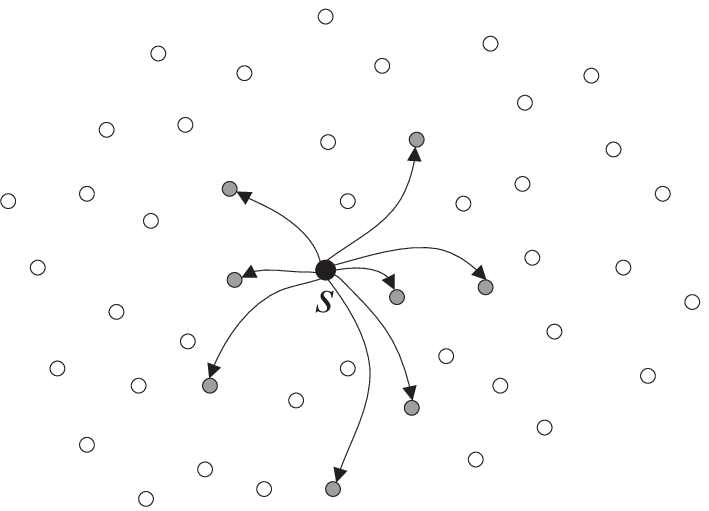}}
  \caption{The social group members (marked with shaded circles) of a source node $s$ when $q=8$.}
  \label{Fig:SocialGrouping}
\end{figure}

In prior studies on the large-sale network analysis,
each source node selects its destination node in a random fashion
without taking into account the geographic distance between the
two nodes, i.e., S--D pairings are {\em randomly} picked so that
each node is the destination of exactly one source. It is thus
likely that the source is far away from its destination. On the
other hand, in our work, nodes are assumed to be {\em
geographically} related with each other in a social context, which
is more feasible in practice. In online social networks, it was
observed that the probability of friendship formation (or social
relationships) between two nodes (users) is proportional to the
inverse of the power of the Euclidean distance between the two
nodes
(users)~\cite{BackstromSunMarlow:WWW10,Liben-NowellNovakKumarRaghavanTomkins:05}.
Each node is allowed to have multiple friends by forming a social
group for the node. A source node and its destination node are chosen only out of its social group members by the central processor that is aware of one-to-one friendship relations. In our distance-based social
formation model, we assume that the Euclidean distance between a
source and its social group members follows a power-law
distribution as
in~\cite{BackstromSunMarlow:WWW10,Liben-NowellNovakKumarRaghavanTomkins:05}.
As illustrated in Fig.~\ref{Fig:SocialGrouping}, a source node $s$
selects other node $v_i$ as its contact with a probability
proportional to $d_{s,v_i}^{-\gamma}$, where $\gamma>0$ is the
social group density. Each node has a social group that consists
of $q$ contacts selected independently, where $q\in
\{1,\cdots,n-1\}$ (i.e., $q=O(n)$). The probability that the set
of nodes $\{v_{i_1},\ldots,v_{i_q}\}$ forms a social group of the
node $s$ is given by
\begin{align}\label{EQ:Prob-SocialGrou}
  \Pr\{\textrm{Social group of~} s=\{v_{i_1},\ldots,v_{i_q}\}\}=\frac{d_{s,v_{i_1}}^{-\gamma}\cdots
  d_{s,v_{i_q}}^{-\gamma}}{N_{\gamma,q}},
\end{align}
where $N_{\gamma,q}=\sum_{1\leq i_1<\cdots<i_q\leq
n}d_{s,v_{i_1}}^{-\gamma}\cdots d_{s,v_{i_q}}^{-\gamma}$ is the
normalization factor. From~\eqref{EQ:Prob-SocialGrou}, one can see
that when the social group density $\gamma$ is large, the social
group members of a node tend to be close to each other. After
forming social groups, each source selects one of its social group
members as its destination uniformly at random.


\subsection{Preliminaries}

In this subsection, we present important lemmas to analyze the
throughput--delay trade-off for the MH and network-decomposed HC
protocols in the subsequent sections. In pure wireless ad hoc
networks with no social group, since the sources and destinations
are paired up one-to-one in a random fashion, the average distance
between an S--D pair is given by $O(1)$ with high probability
(w.h.p.) in a dense network. On the other hand, in the context of
social relationships, the average distance of an S--D pair depends
on both the number of contacts in a social group, $q$, and the
social group density $\gamma$ since a source node selects one of
its social contacts in the belonging social group as its
destination node randomly. The average distance between an S--D
pair is specified according to the parameters $q$ and $\gamma$
below.

\begin{lemma}\label{Lem:AverageDistance}
If $q$ scales faster than a constant independent of $n$ (i.e.,
$q=\omega(1)$), then the average distance between a source $s$ and
its destination $v$, $\mathbb{E}[d_{s,v}]$, in a dense network is
given by
\begin{align}
  \mathbb{E}[d_{s,v}]= \Theta(1). \label{EQ:Edsv1}
\end{align}
If $q=\Theta(1)$, then $\mathbb{E}[d_{s,v}]$ in a dense network is
given by
\begin{align} 
    \mathbb{E}[d_{s,v}]=\left\{
    \begin{array}{ll}
      \Theta\left(1\right) & \textrm{for~} 0\leq\gamma<2 \\
      \Theta\left(\left(\frac{\log n}{n}\right)^{\frac{\gamma}{2}-1}\right) & \textrm{for~} 2\leq\gamma\leq 3 \\
      \Theta\left(\sqrt{\frac{\log n}{n}}\right) & \textrm{for~}
      \gamma>3,
    \end{array}
    \right. \label{EQ:Edsv2}
\end{align}
where $\gamma>0$ denotes the social group density.
\end{lemma}

\begin{IEEEproof}
From~\cite[Section
IV]{AzimdoostSadjadpourGarcia-Luna-Aceves:TWC13}, when the MH
protocol is employed in a dense network, the average number of
hops in any given S--D routing path, $\mathbb{E}[X]$, is shown to
be
\begin{align}
\mathbb{E}[X]= \left\{
\begin{array}{ll}
  \Theta\left(\frac{1}{r(n)}\right) & \textrm{for~} q=\omega(1) \\
  \Theta\left(\frac{n}{n-q+1}\frac{1}{r(n)}
  \right) & \textrm{for~} q=\Theta(1), 0\leq\gamma<2 \\
  \Theta\left(\frac{n}{n-q+1}\frac{1}{r^{3-\gamma}(n)}\right)
  & \textrm{for~} q=\Theta(1), 2\leq\gamma\leq 3 \\
  \Theta\left(\frac{n}{n-q+1}\right)
  & \textrm{for~} q=\Theta(1), \gamma>3,
\end{array}
\right. \nonumber
\end{align}
where $r(n)=\Theta\left(\sqrt{\frac{\log n}{n}}\right)$ is the
transmission range per hop. By multiplying $\mathbb{E}[X]$ by
$r(n)$, the average distance $\mathbb{E}[d_{s,v}]$ can be
expressed as
\begin{align}
    \mathbb{E}[d_{s,v}]
    =\left\{
    \begin{array}{ll}
      \Theta\left(1\right) & \textrm{for~} q=\omega(1) \\
      \Theta\left(\frac{n}{n-q+1}\right) & \textrm{for~} q=\Theta(1), 0\leq\gamma<2 \\
      \Theta\left(\frac{n}{n-q+1}\left(\frac{\log n}{n}\right)^{\frac{\gamma}{2}-1}\right) & \textrm{for~} q=\Theta(1), 2\leq\gamma\leq 3 \\
      \Theta\left(\frac{n}{n-q+1}\sqrt{\frac{\log n}{n}}\right) & \textrm{for~} q=\Theta(1),
      \gamma>3,
    \end{array}
    \right. \nonumber
\end{align}
which thus leads to (\ref{EQ:Edsv1}) and (\ref{EQ:Edsv2}). This
completes the proof of this lemma.
\end{IEEEproof}

Since the distance between two nodes in an extended network is
simply increased by a factor of $\sqrt{n}$ compared to a dense
network, the average distance in the extended network is also
increased by a factor of $\sqrt{n}$.

A natural question arises to examine whether the distance between
an S--D pair may deviate from its mean. In the following lemma, we
show that the distance of an S--D pair does not scale faster than
the average distance $\mathbb{E}[d_{s,v}]$ w.h.p.

\begin{lemma}\label{Lem:AverageDistanceMarkov}
The distance between a source $s$ and its destination $v$,
$d_{s,v}$, does not scale at a faster rate than the average
distance $\mathbb{E}[d_{s,v}]$ within a factor of $n^\epsilon$
w.h.p. for an arbitrarily small $\epsilon>0$. That is, the
probability $\Pr\{d_{s,v}< n^{\epsilon}\mathbb{E}[d_{s,v}]\}$ is
given by $1-\frac{1}{n^{\epsilon}}$.
\end{lemma}

\begin{IEEEproof}
By Markov inequality,\footnote{One can use a tighter inequality
with the second moment of random variable $d_{s,v}$ (e.g.,
Chebyshev's inequality), which however does not fundamentally
change our main results.} we have
\begin{align}
  \Pr\{d_{s,v}< n^{\epsilon}\mathbb{E}[d_{s,v}]\}
  &=1-\Pr\{d_{s,v}\geq n^{\epsilon}\mathbb{E}[d_{s,v}]\}
  \nonumber\\
  &\geq 1- \frac{\mathbb{E}[d_{s,v}]}{n^{\epsilon}\mathbb{E}[d_{s,v}]}
  =1-\frac{1}{n^{\epsilon}}, \nonumber
\end{align}
which tends to one as $n$ goes to infinity. This completes the
proof of this lemma.
\end{IEEEproof}

Hence from Lemmas~\ref{Lem:AverageDistance} and
\ref{Lem:AverageDistanceMarkov}, one can straightforwardly replace
the average distance $\mathbb{E}[d_{s,v}]$ by the distance
$d_{s,v}$ of a certain S--D pair as long as a factor of
$n^\epsilon$ can be ignored.

\subsection{Performance Metrics}

In this subsection, we formally define the throughput and delay
used throughout the paper.

\begin{definition}[Throughput]
A per-node throughput $R(n)$ is said to be {\em achievable} w.h.p.
if all sources can transmit at the average rate of $R(n)$
bits/s/Hz to their destinations with probability approaching one
as $n$ increases. The achievable aggregate throughput is given by
$T(n)=\Omega(nR(n))$ accordingly.
\end{definition}

%


\begin{definition}[Delay]
The end-to-end delay of a packet is the time that it takes for the
packet to reach its destination after it leaves the
source. The delay $D(n)$ is the expectation of
the average delay over all S--D pairs and is expressed as
\begin{align}
  D(n)=\frac{2}{n}\sum_{i=1}^{n/2}D_i(n),
  \nonumber
\end{align}
where $D_i(n)$ is the sample mean of delay (over packets that reach their destinations) for S--D pair $i$.
\end{definition}

Note that the queuing delay at the source node is
not included in our work. Although the queuing delay may account for a large portion of the overall delay of a packet, it was shown in~\cite{ElGamalMammenPrabhakarShah:TIT06-2} that when the queuing delay is taken into account, the throughput--delay trade-off for the MH protocol remains unchanged in order sense since the average delay at each server, corresponding to a routing cell, is bounded by some constant independent of $n$. Hence, Theorems~\ref{THM:TD-Tradeoff-MH-Dense} and \ref{THM:TD-Tradeoff-MH-Extended} in Section~\ref{SEC:MainResults} will remain the same even with the queuing delay. On the other hand, for the HC protocol, the analysis of delay scaling with the queuing delay is left open even under network models without social relationships.

\section{Main Results}\label{SEC:MainResults}

In this section, we present main results of this paper by
characterizing the throughput--delay trade-off achieved by the MH
and network-decomposed HC protocols with the distance-based social
formation model in Section~\ref{SEC:SocialModel}. The
throughput--delay trade-off depends on the path-loss exponent
$\alpha$ and the parameters of the social formation model such as
the social group density $\gamma$ and the number of contacts in a
social group, $q$. To illustrate the main results more concisely,
we will first identify operating regimes on the throughput--delay
trade-off as follows:
\begin{itemize}
  \item Regime A (Low social group density regime): $q=\omega(1)$ or $\{q=\Theta(1)~\text{and}~0\leq\gamma<2\}$
  \item Regime B (Medium social group density regime): $q=\Theta(1)$ and $2\leq\gamma\leq 3$
  \item Regime C (High social group density regime): $q=\Theta(1)$ and $\gamma>3$.
\end{itemize}
That is, the entire operating regimes are categorized by $\gamma$
and $q$.

Note that similarly as in other studies on the
throughput scaling laws of ad hoc networks
\cite{OzgurLeveque:TIT10,ElGamalMammenPrabhakarShah:TIT06-2,ElGamalMammenPrabhakarShah:TIT06,NeelyModiano:05,GuptaKumar:00,FranceschettiDouseTseThiran:07,GuptaKumar:03,XueXieKumar:05,ShinChungLee:TIT13,OzgurLevequeTse:07,NiesenGuptaShah:09,ZemlianovVeciana:05,O.Dousse:INFOCOM02,ShinJeonDevroyeVuChungLeeTarokh:08,GrossglauserTse:02,ZhangXuWangGuizani:TC10,LiZhangFang:TMC11,YoonShinJeon:TMC17,LeeChung:12,
NebatCruzBhardwaj:09}, it is possible to have no outage event by
globally controlling interference based on the TDMA operation
among square routing cells. Thus, we do not employ any
retransmission scheme in our case since we will make the outage
probability approach zero. Next, we will briefly account for the
MH protocol in ad hoc networks, which was originally introduced
in~\cite{GuptaKumar:00} and then was generalized
in~\cite{ElGamalMammenPrabhakarShah:TIT06}. The overall procedure
of the MH protocol is described as follows:
\begin{itemize}
  \item The network is divided into square routing cells of area $a(n)=\Omega(\log n/n)$ in a dense network ($a(n)=\Omega(\log n)$ in an extended network) ensuring that each cell includes at least one node w.h.p.~\cite{GuptaKumar:00,ElGamalMammenPrabhakarShah:TIT06}.
  \item Draw a line connecting a source to its destination and perform MH routing horizontally or vertically by using the adjacent routing cells passing through the line until its packet reaches the corresponding destination.
  \item By virtue of the central processor, each routing cell operates the $t$-TDMA to avoid huge interference, where $t > 0$ is some small constant independent of $n$.
\end{itemize}
The network-decomposed HC protocol will be described in detail in
Section~\ref{SEC:NDHC}.

\subsection{Throughput--Delay Trade-off in Dense Networks}

In this subsection, we show the throughput--delay trade-off
achieved by the MH and network-decomposed HC protocols in a dense
network. The number of S--D lines passing through each cell in the
MH protocol is first specified in the following lemma.

\begin{lemma}\label{Lem:NumSDpairs}
For $a(n)=\Omega(\log n/n)$, the total number of S--D lines
passing through each cell is $O(\mathbb{E}[d_{s,v}]n\sqrt{a(n)})$, where $\mathbb{E}[d_{s,v}]$
is the average distance of an S--D pair.
\end{lemma}
\begin{IEEEproof}
By generalizing the arguments
in~\cite[Lemma~3]{ElGamalMammenPrabhakarShah:TIT06} to the case of
an arbitrary average distance of an S--D pair, one can easily show
that the total number of S--D lines passing through each cell is
given by $O(\mathbb{E}[d_{s,v}]n\sqrt{a(n)})$. This completes the
proof of this lemma.
\end{IEEEproof}

The main result achieved by the MH protocol in a dense network is
shown in the following theorem.

\begin{theorem}\label{THM:TD-Tradeoff-MH-Dense}
The throughput--delay trade-off for the MH protocol in a wireless
{\it dense} network adopting the social formation model is given
by
\begin{align} \label{EQ:TD_dense}
(T(n),D(n))
=\left\{
\begin{array}{ll}
  \left(\Theta\left(\frac{1}{\sqrt{a(n)}}\right),\Theta\left(\frac{1}{\sqrt{a(n)}}\right)\right)
  \\
  \textrm{for Regime~A}\\
  \left(\Theta\left(\left(\frac{n}{\log n}\right)^{\frac{\gamma}{2}-1}\frac{1}{\sqrt{a(n)}}\right)
    ,\Theta\left(\left(\frac{\log n}{n}\right)^{\frac{\gamma}{2}-1}\frac{1}{\sqrt{a(n)}}\right)\right)
  \\
  \textrm{for Regime~B}\\
  \left(\Theta\left(\sqrt\frac{n}{a(n)\log n}\right)
    ,\Theta\left(\sqrt\frac{\log n}{a(n)n}\right)\right)
  \\
  \textrm{for Regime~C},
\end{array}
\right.
\end{align}
where the area of each square cell is given by $a(n)=\Omega(\log
n/n)$.
\end{theorem}

\begin{IEEEproof}
Using Lemmas~\ref{Lem:AverageDistance} and~\ref{Lem:NumSDpairs},
the number of S--D lines passing through each cell, $u(n)$, is
given by
\begin{align}
u(n)=\left\{
\begin{array}{ll}
  O\left(n\sqrt{a(n)}\right) & \textrm{for Regime~A}\\
  O\left(\left(\frac{\log n}{n}\right)^{\frac{\gamma}{2}-1}n\sqrt{a(n)}\right) & \textrm{for Regime~B}\\
  O\left(\sqrt{\frac{\log n}{n}}n\sqrt{a(n)}\right) & \textrm{for Regime~C}.
\end{array}
\right. \nonumber
\end{align}
Due to the fact that the total traffic through each cell is the
traffic generated by all the S--D lines passing through the cell
and the cell throughput is $\Theta(1)$ under the TDMA operation,
the aggregate throughput $T(n)$ is lower-bounded by
\begin{align} \label{EQ:Tn_lower}
T(n)&=\Theta\left(\frac{n}{u(n)}\right)
\nonumber\\
&=\left\{
\begin{array}{ll}
  \Theta\left(\frac{1}{\sqrt{a(n)}}\right) & \textrm{for Regime~A}\\
  \Theta\left(\left(\frac{n}{\log n}\right)^{\frac{\gamma}{2}-1}\frac{1}{\sqrt{a(n)}}\right) & \textrm{for Regime~B}\\
  \Theta\left(\sqrt\frac{n}{\log n}\frac{1}{\sqrt{a(n)}}\right) & \textrm{for
  Regime~C}.
\end{array}
\right.
\end{align}

We now turn to computing the average packet delay. Since each hop
covers a distance of $\Theta(\sqrt{a(n)})$ and the delay is at
most a constant times the number of hops, from
Lemma~\ref{Lem:AverageDistance}, the delay $D(n)$ is given by
\begin{align} \label{EQ:Hn}
D(n)&=\Theta\left(\frac{\mathbb{E}[d_{s,v}]}{\sqrt{a(n)}}\right)
\nonumber\\
&=\left\{
\begin{array}{ll}
  \Theta\left(\frac{1}{\sqrt{a(n)}}\right) & \textrm{for Regime~A}\\
  \Theta\left(\left(\frac{\log n}{n}\right)^{\frac{\gamma}{2}-1}\frac{1}{\sqrt{a(n)}}\right) & \textrm{for Regime~B}\\
  \Theta\left(\sqrt\frac{\log n}{n}\frac{1}{\sqrt{a(n)}}\right) & \textrm{for
  Regime~C}.
\end{array}
\right.
\end{align}
Hence, using (\ref{EQ:Tn_lower}) and (\ref{EQ:Hn}) finally leads
to (\ref{EQ:TD_dense}), which completes the proof of this theorem.
\end{IEEEproof}

From the above result, the following observations can be found
according to each operating regime. In Regime A (i.e., the low
social group density regime), the throughput--delay trade-off for
the MH protocol is the same as that
in~\cite{ElGamalMammenPrabhakarShah:TIT06}, in which no social
group exists. This is because small values of $\gamma$ correspond
to the case where the social group members are widely distributed
over the whole network and hence the average distance of an S--D
pair is not reduced compared to the network case with no social
group. In Regime B, as $\gamma$ increases, the throughput is
improved and the delay is reduced by virtue of the decreased
average distance of an S--D pair. In Regime C, the maximum
throughput $\Theta(n)$ can be achieved by only using $\Theta(1)$
hops. This is consistent with the throughput scaling result
according to the social group density
in~\cite{AzimdoostSadjadpourGarcia-Luna-Aceves:TWC13}.

\begin{figure}[t!]
  \centering
  \includegraphics[width=3.3in]{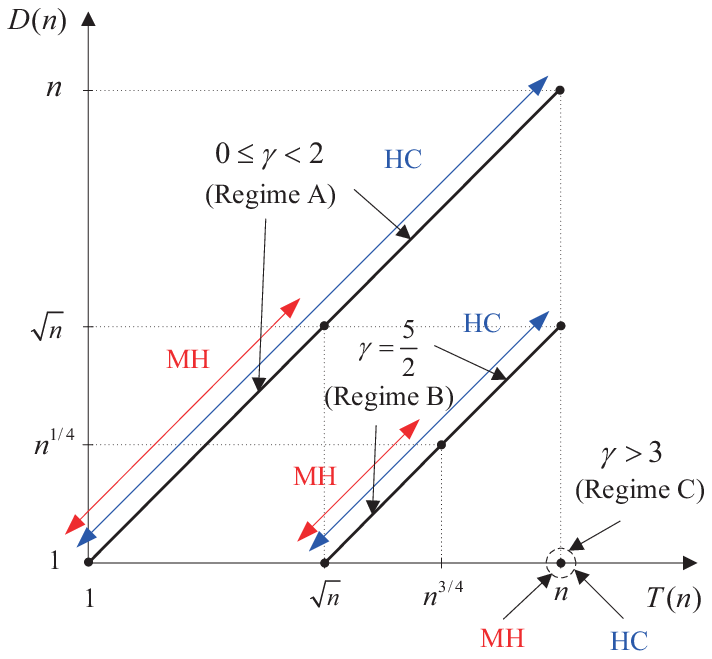}\\
  \caption{The throughput--delay trade-off in a dense network when $q=\Theta(1)$. Red and blue arrows denote the trade-offs for the MH and network-decomposed HC protocols, respectively. Black lines indicate the best trade-offs achieved by one of these two protocols. The factor of $\log n$ is omitted for simplicity.}
  \label{Fig:TDtradeoff-Dense}
\end{figure}

\begin{remark}\label{Remark:TD-tradeoff-MH}
In our dense network, the throughput--delay trade-off for the MH
protocol is illustrated in Fig.~\ref{Fig:TDtradeoff-Dense} when
$q=\Theta(1)$. The red arrows in Fig.~\ref{Fig:TDtradeoff-Dense}
represent the throughput--delay scaling results achieved by the MH
protocol according to different values of $\gamma$. More
specifically, when $0\leq\gamma<2$ (Regime A), the trade-off is
given by
$(T(n),D(n))=\left(\frac{1}{\sqrt{a(n)}},\frac{1}{\sqrt{a(n)}}\right)$
for $a(n)=\Omega\left(\frac{\log n}{n}\right)$ and $a(n)=O(1)$. In
Regime B, as an exemplary value of $\gamma$, the trade-off for
$\gamma=\frac{5}{2}$ is illustrated in the figure for
$a(n)=\Omega\left(\frac{\log n}{n}\right)$ and
$a(n)=O\left(\frac{1}{\sqrt{n}}\right)$, where
$(T(n),D(n))=\left(\Theta\left(\left(\frac{n}{\log
n}\right)^{\frac{1}{4}}\frac{1}{\sqrt{a(n)}}\right)
,\Theta\left(\left(\frac{\log
n}{n}\right)^{\frac{1}{4}}\frac{1}{\sqrt{a(n)}}\right)\right)$.
When $\gamma>3$ (Regime C), the trade-off is given by
$(T(n),D(n))=\left(\Theta\left(\frac{n}{\log n}\right)
,\Theta(1)\right)$ for $a(n)=\Theta\left(\frac{\log n}{n}\right)$,
which corresponds to a single point in the figure. As seen in
Fig.~\ref{Fig:TDtradeoff-Dense}, the throughput--delay trade-off
gets improved significantly as the social group density $\gamma$
 increases.
\end{remark}

\begin{remark}\label{Remark:TD-tradeoff-MH-Other}
To better understand the above trade-off between throughput and
delay for the MH protocol in the dense network, we express $T(n)$
as a function of $D(n)$ as follows:
\begin{align}\label{TD-Relationship-MH}
T(n)=\left\{
\begin{array}{ll}
  \Theta\left(D(n)\right) & \textrm{for Regime~A}\\
  \Theta\left(D(n)n^{\gamma-2-\epsilon}\right) & \textrm{for Regime~B}\\
  \Theta\left(D(n)n^{1-\epsilon}\right) & \textrm{for Regime~C}
\end{array}
\right.
\end{align}
for an arbitrarily small $\epsilon>0$. It is again observed that
the throughput $T(n)$ given the delay $D(n)$ can be greatly
improved as $\gamma$ increases.
\end{remark}

We modify the HC protocol in~\cite{OzgurLeveque:TIT10} so that the
network under our social formation model operates properly, which
will be explained in detail in Section~\ref{SEC:NDHC}. We herein
summarize our main results for the proposed network-decomposed HC
protocol when the notion of social relationships is incorporated
into ad hoc networks. The main results for a dense network are
shown as follows.

\begin{theorem}\label{THM:TD-Tradeoff-HC}
The throughput--delay trade-off for the network-decomposed HC
protocol in a wireless {\it dense} network adopting the social
formation model is given by
\begin{align}\label{TD-Tradeoff-HC-Dense}
(T(n),D(n))
= \left\{
\begin{array}{ll}
  \left(\Theta\left(n^{b-\epsilon}\right), \Theta\left(n^{b+\epsilon}\right)\right)  
  &\textrm{for Regime~A}\\
  \left(\Theta\left(n^{b-(\gamma-2)(b-1)-\epsilon}\right),\Theta\left(n^{(3-\gamma)b+\epsilon}\right)\right)
  &\textrm{for Regime~B}\\
  \left(\Theta\left(n^{1-\epsilon}\right),\Theta\left(n^{\epsilon}\right)\right)  
  &\textrm{for Regime~C}
\end{array}
\right.
\end{align}
for an arbitrarily small $\epsilon>0$, where $0\leq b<1$.
\end{theorem}
\begin{IEEEproof}
See Appendix~\ref{SEC:Proof-1}.
\end{IEEEproof}

From Theorem~\ref{THM:TD-Tradeoff-HC}, the following insightful
observations are found according to each operating regime. In
Regime A, it is shown that the throughput--delay trade-off derived
along with the social formation model is the same as that with no
social behavior. In other words, the network behaves as if there
is no social group when the social group density $\gamma$ is
small. In Regime B, as $\gamma$ increases, the throughput is
improved and the delay is reduced. In Regime C, the maximum
throughput $\Theta(n)$ is achieved with a very small delay. These
observations are similar to those for the MH case. The
throughput--delay trade-off for the network-decomposed HC protocol
is illustrated in Fig.~\ref{Fig:TDtradeoff-Dense} (see blue
arrows).

\begin{remark}
When $0<\gamma<2$ (Regime A), the throughput--delay trade-off for
the network-decomposed HC protocol is given by
$(T(n),D(n))=(\Theta(n^{b-\epsilon}),\Theta(n^{b+\epsilon}))$ for
$0\leq b<1$. In Regime B, the trade-off achieved by the
network-decomposed HC protocol for $\gamma=\frac{5}{2}$ (as an
exemplary value of $\gamma$) is given by
$(T(n),D(n))=(\Theta(n^{(b+1)/2-\epsilon}),\Theta(n^{b/2+\epsilon}))$
for $0\leq b<1$. When $\gamma>3$ (Regime C), the trade-off for the
network-decomposed HC protocol is given by
$(T(n),D(n))=(\Theta(n^{1-\epsilon}),\Theta(n^{\epsilon}))$. From
Fig.~\ref{Fig:TDtradeoff-Dense}, it is seen that for $0\leq b\leq
\frac{1}{2}$, performance on the trade-off for the MH and
network-decomposed HC protocols is the same, but for $\frac{1}{2}
< b < 1$, the network-decomposed HC protocol has a higher
throughput than that of the MH protocol at the expense of an
increased delay. In the dense network, the best trade-offs can
thus be achieved by the network-decomposed HC protocol and are
depicted by black lines in Fig.~\ref{Fig:TDtradeoff-Dense}.
\end{remark}

\begin{remark}
To better understand the above trade-off between throughput and
delay for the network-decomposed HC protocol in the dense network,
we express $T(n)$ as a function of $D(n)$ as follows:
\begin{align}\label{TD-Relationship-HC-Dense}
T(n)=\left\{
\begin{array}{ll}
  \Theta\left(D(n)n^{-\epsilon}\right) & \textrm{for Regime~A}\\
  \Theta\left(D(n)n^{\gamma-2-\epsilon}\right) & \textrm{for Regime~B}\\
  \Theta\left(D(n)n^{1-\epsilon}\right) & \textrm{for Regime~C}
\end{array}
\right.
\end{align}
for an arbitrarily small $\epsilon>0$. It is observed that the
throughput--delay trade-off in~\eqref{TD-Relationship-HC-Dense}
for the network-decomposed HC protocol is the same as
\eqref{TD-Relationship-MH} for the MH protocol within a factor of
$n^\epsilon$ as long as the delay $D(n)$ scales up to its maximum
achieved by MH (refer to Fig.~\ref{Fig:TDtradeoff-Dense}). It
means that the inherent relation between the two protocols with
respect to the throughput--delay trade-off remains the same, but a
general trade-off can be achieved by incorporating the results for
the network-decomposed HC protocol.
\end{remark}

\subsection{Throughput--Delay Trade-off in Extended Networks}\label{SEC:TD_HC}

We next show the throughput--delay trade-off of an extended
network. The trade-off achieved by the MH protocol is first
presented in the following theorem.

\begin{theorem}\label{THM:TD-Tradeoff-MH-Extended}
The throughput--delay trade-off for the MH protocol in a wireless
{\it extended} network adopting the social formation model is
given by
\begin{align}\label{EQ:TD-Tradeoff-MH-Extended}
(T(n),D(n))
=\left\{
\begin{array}{ll}
  \left(\Theta\left(\sqrt{\frac{n}{a(n)^{\alpha+1}}}\right)
  ,\Theta\left(\sqrt{\frac{n}{a(n)}}\right)\right)  
  &\textrm{for Regime~A}\\
  \left(\Theta\left(\left(\frac{n}{\log n}\right)^{\frac{\gamma}{2}-1}\sqrt{\frac{n}{a(n)^{\alpha+1}}}\right),  
  \Theta\left(\left(\frac{\log n}{n}\right)^{\frac{\gamma}{2}-1}\sqrt{\frac{n}{a(n)}}\right)\right)  
  &\textrm{for Regime~B}\\
  \left(\Theta\left(\frac{n}{\sqrt{a(n)^{\alpha+1}\log n}}\right)
    ,\Theta\left(\sqrt\frac{\log n}{a(n)}\right)\right)  
  &\textrm{for Regime~C}
\end{array}
\right.
\end{align}
where the area of each square cell is given by $a(n)=\Omega(\log
n)$.
\end{theorem}

\begin{IEEEproof}
Similarly as in the proof of
Theorem~\ref{THM:TD-Tradeoff-MH-Dense}, the throughput--delay
trade-off for the MH protocol in an extended network can easily be
proved by computing the received signal power that is expressed in
a different manner due to the power limitation as well as using
the fact that the area of each cell, $a(n)$, is increased by a
factor of $n$ compared to a dense network. Since the received
signal-to-noise ratio (SNR) for the signal transmitted from the
nearest-neighbor cell scales as
$\left(\frac{1}{\sqrt{a(n)}}\right)^\alpha$, we obtain the
throughput result in~\eqref{EQ:TD-Tradeoff-MH-Extended} by using
(\ref{EQ:TD_dense}), which completes the proof of this theorem.
\end{IEEEproof}

\begin{figure}[t!]
  \centering
  \subfigure[$\alpha=2$]{
  \includegraphics[width=3.3in]{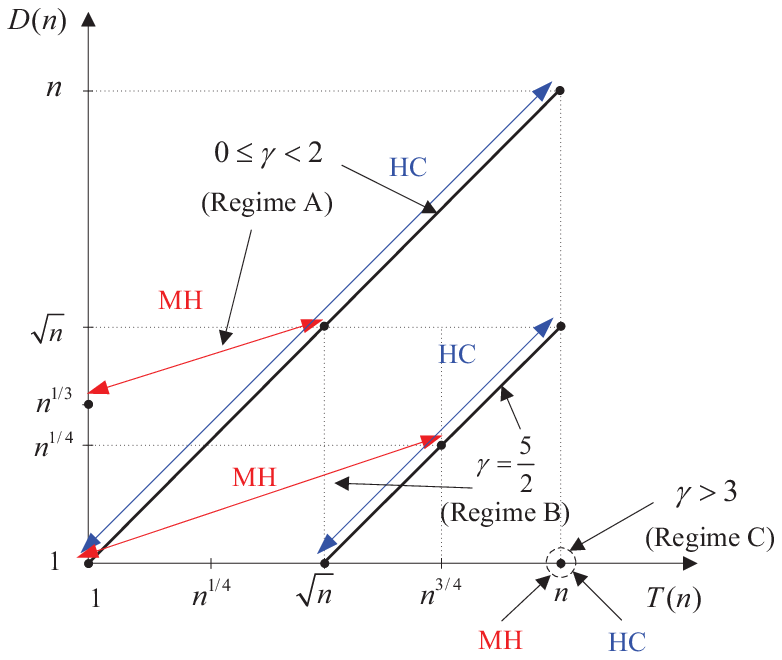}}
  \hspace{1.0cm}
  \centering
  \subfigure[$\alpha=3$]{
  \includegraphics[width=3.3in]{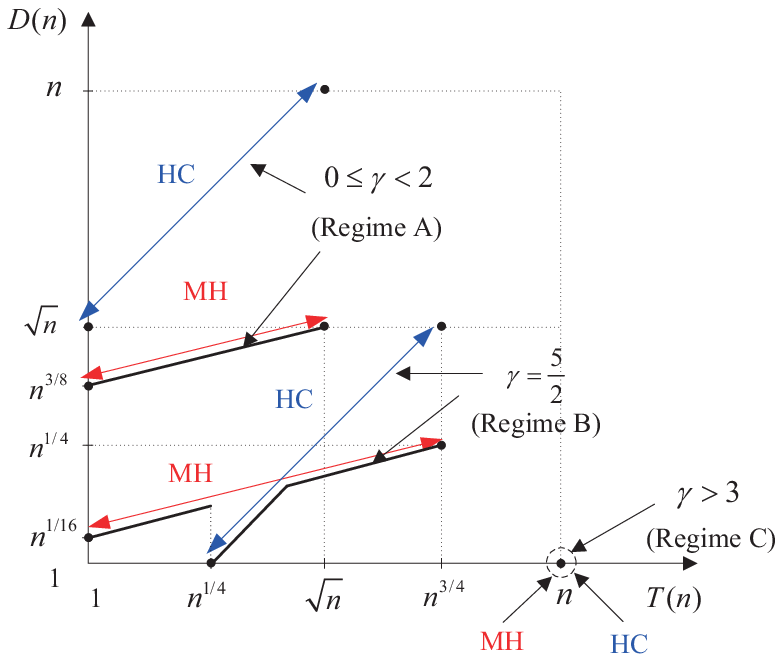}}
  \caption{The throughput--delay trade-off in an extended network when $q=\Theta(1)$. Red and blue arrows denote the trade-offs for the MH and network-decomposed HC protocols, respectively. Black lines indicate the best trade-offs achieved by one of these two protocols. The factor of $n^\epsilon$ is omitted for simplicity.}
  \label{Fig:TDtradeoff-Extended}
\end{figure}

In the extended network, the throughput--delay trade-off for the
MH protocol is illustrated in Fig.~\ref{Fig:TDtradeoff-Extended}
when $q=\Theta(1)$ and $\alpha\in\{2,3\}$. The red arrows in the
figure represent the throughput--delay scaling results achieved by
the MH protocol according to different values of $\gamma$.

\begin{remark}\label{Remark:TD-tradeoff-MH-Extended}
To better understand the above trade-off between throughput and
delay for the MH protocol in the extended network, we express
$T(n)$ as a function of $D(n)$ as follows:
\begin{align} \nonumber
T(n)=\left\{
\begin{array}{ll}
  \Theta\left(D(n)a(n)^{-\alpha/2}\right) & \textrm{for Regime~A}\\
  \Theta\left(D(n)n^{\gamma-2-\epsilon}a(n)^{-\alpha/2}\right) & \textrm{for Regime~B}\\
  \Theta\left(D(n)n^{1-\epsilon}a(n)^{-\alpha/2}\right) & \textrm{for Regime~C}
\end{array}
\right.
\end{align}
for an arbitrarily small $\epsilon>0$. Compared to the dense
network case (see Remark~\ref{Remark:TD-tradeoff-MH-Other}), the
throughput $T(n)$ given the delay $D(n)$ is reduced by a factor of
$a(n)^{-\alpha/2}$, which corresponds to the amount of SNR loss.
\end{remark}

Now, let us turn to summarizing our main results for an extended
network when the proposed network-decomposed HC protocol is
employed.

\begin{theorem}\label{THM:TD-Tradeoff-HC-Extended}
The throughput--delay trade-off for the network-decomposed HC
protocol in a wireless {\it extended} network adopting the social
formation model is given by
\begin{align} \label{EQ:TD-Tradeoff-HC-Extended}
(T(n),D(n)) = \left\{
\begin{array}{ll}
  \left(\Theta\left(n^{b-\alpha/2+1-\epsilon}\right), \Theta\left(n^{b+\epsilon}\right)\right)  
  &\textrm{for Regime~A}\\
  \left(\Theta\left(n^{(b-\alpha/2)(3-\gamma)+1-\epsilon}\right),\Theta\left(n^{(3-\gamma)b+\epsilon}\right)\right)  
  &\textrm{for Regime~B}\\
  \left(\Theta\left(n^{1-\epsilon}\right),\Theta\left(n^{\epsilon}\right)\right)  
  &\textrm{for Regime~C}
\end{array}
\right.
\end{align}
for an arbitrarily small $\epsilon>0$, where $0\leq b<1$.
\end{theorem}
\begin{IEEEproof}
See Appendix~\ref{SEC:Proof-2}.
\end{IEEEproof}

In contrast to the results for the dense network in
\eqref{TD-Tradeoff-HC-Dense}, the throughput scaling achieved by
the network-decomposed HC protocol in the extended network depends
highly on the path-loss exponent $\alpha$ in Regimes A and B
(i.e., the low and medium social group density regimes,
respectively), while the delay scaling is the same for both dense
and extended networks. This is because in the above two regimes, a
bursty transmission whose duration depends on $\alpha$ is needed
due to the power limitation, which will be explained in detail in
Section~\ref{SEC:ExtendedNetwork}.

\begin{remark}
From Theorem~\ref{THM:TD-Tradeoff-HC-Extended}, $T(n)$ for the
network-decomposed HC ptorocol in an extended network is expressed
as a function of $D(n)$ as follows:
\begin{align} \nonumber
T(n)
=\left\{
\begin{array}{ll}
  \Theta\left(D(n)n^{-\alpha/2+1-\epsilon}\right) & \textrm{for Regime~A}\\
  \Theta\left(D(n)n^{1+\alpha(\gamma-3)/2-\epsilon}\right) & \textrm{for Regime~B}\\
  \Theta\left(D(n)n^{1-\epsilon}\right) & \textrm{for Regime~C}
\end{array}
\right.
\end{align}
for an arbitrarily small $\epsilon>0$.
\end{remark}

In the extended network, the throughput--delay trade-off for the
network-decomposed HC protocol is also illustrated in
Fig.~\ref{Fig:TDtradeoff-Extended} when $q=\Theta(1)$ and
$\alpha\in\{2,3\}$ (see blue arrows), and the following
interesting observations in comparison with the results for the MH
protocol are made according to each operating regime.

\begin{remark}
First, it is shown in Fig.~\ref{Fig:TDtradeoff-Extended} that the
trade-offs for the MH and network-decomposed HC protocols are
improved as the social group density $\gamma$ increases as in the
dense network configuration. We now recall that in the extended
network, the throughput for a given delay depends on the path-loss
exponent $\alpha$, unlike the dense network case. When $\alpha=2$,
the trade-off achieved by the network-decomposed HC protocol is
always superior to or equal to that of the MH protocol for all the
operating regimes. On the other hand, as $\alpha$ increases, there
exists operating regimes such that the MH protocol is dominant.
More precisely, in Regime A, it is not difficult to show that if
$\alpha>1+\sqrt{3}$, then the MH protocol outperforms the
network-decomposed HC protocol, and there is a crossover between
two lines achieved by both protocols otherwise. In Regime B, the
MH protocol is superior to the network-decomposed HC protocol if
\begin{align}
  \gamma\le \frac{3\alpha^2-2\alpha-6}{\alpha^2-2}. \nonumber
\end{align}
Otherwise, there is a crossover between two lines achieved by both
protocols, where the network-decomposed HC protocol has better
trade-off performance in the low throughput regime as depicted in
Fig.~\ref{Fig:TDtradeoff-Extended}(b). In Regime C, both MH and
network-decomposed HC protocols achieve the best throughput of
$\Theta(n)$ with a very small delay of $\Theta(1)$ in which the
social group density $\gamma$ becomes high, as in the dense
network case.
\end{remark}

%

\section{Throughput--Delay Trade-off for Network-Decomposed HC}\label{SEC:TD-Tradeoff-HC}

In this section, we shall first introduce the network-decomposed
HC protocol so that the network under our social formation model
operates suitably, and then derive its throughput--delay trade-off
in a dense network. The trade-off achieved by the
network-decomposed HC protocol in an extended network is also
shown through network transformation.

\subsection{Conventional HC Protocol}

We will first briefly explain the original HC
protocol~\cite{OzgurLevequeTse:07} and its modified
one~\cite{OzgurLeveque:TIT10} in an ad hoc network. It will be
shown how the HC protocol is further modified to work properly for
our network with the distance-based social formation model in the
subsequent subsections. In the original HC
protocol~\cite{OzgurLevequeTse:07}, packets are transmitted
through the following three phases:
\begin{itemize}
  \item The network is divided into multiple clusters, each having $M$
  nodes, where $M=O(n)$.
  \item During the first phase, each source distributes its $M$ bits to other nodes in the same cluster, one bit for each node.
  \item During the second phase, a long-range multiple-input multiple-out (MIMO) transmission is then performed between two clusters having a source and its destination, one at a time.
  \item During the last phase, each node in a cluster quantizes its received observations and delivers the quantized data to the rest of nodes in the same cluster. Each destination can decode its packets by collecting all quantized observations.
\end{itemize}
When each node transmits data within its cluster, another
smaller-scaled cooperation can be applied in the same manner by
dividing each cluster into smaller ones. By recursively applying
this procedure, it is possible to establish the hierarchical
strategy in the network. One drawback of the original HC
protocol~\cite{OzgurLevequeTse:07} is that it requires an
extremely large bulk-size $\Theta(n^{\frac{h}{2}})$, i.e., the
minimum number of bits that should be transmitted between each
S--D pair is large, where $h>0$ is the hierarchy level. Due to a
large bulk-size and inefficient scheduling, the delay scaling of
the HC protocol can be much worse than that of the MH protocol. In
order to improve the delay performance, the original HC protocol
was modified in~\cite{OzgurLeveque:TIT10} by reducing the
bulk-size and enhancing scheduling. Using the modified HC
protocol~\cite{OzgurLeveque:TIT10}, the throughput--delay
trade-off is given by
\begin{align} \nonumber
  (T(n),D(n))=\Theta(n^b/\log n, n^b\log n)
\end{align}
where $0\leq b<1$. The parameter $b$ depends on the size of a
cluster with $0\leq b\leq \frac{h}{h+1}$.

\subsection{Network-Decomposed HC Protocol} \label{SEC:NDHC}

Both the original HC protocol~\cite{OzgurLevequeTse:07} and its
modified one~\cite{OzgurLeveque:TIT10} were designed assuming that
the geographic distance of between a source node and its
destination node scales as $O(1)$ in a dense network due to the
random S--D pairings. Since the distance of an S--D pair in our
network may scale at a lower rate than that of the network size
depending on the parameters of the social formation model, the HC
protocols in~\cite{OzgurLevequeTse:07,OzgurLeveque:TIT10} may not
work effectively. This motivates us to further modify the HC
protocol in~\cite{OzgurLeveque:TIT10} by taking into account the
notion of {\em network decomposition} as in the following.

\begin{figure}[t!]
  \centering
  \includegraphics[width=2.5in]{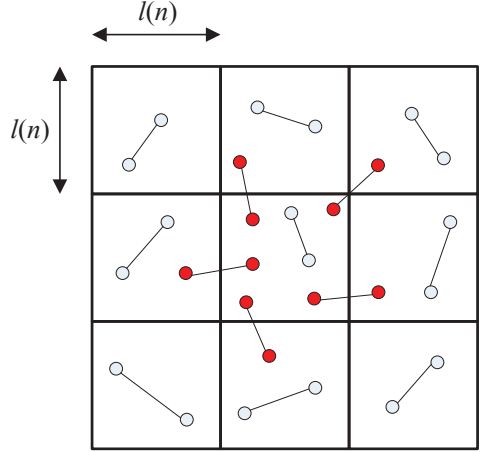}\\
  \caption{The illustration of subnetworks for the network-decomposed HC protocol, where a subnetwork is depicted as a box, a source or destination node is denoted by a circle, and the line between two circles indicates the relationship of S--D pairings. The red circle represents the node whose transmission is performed across the adjacent subnetwork.}
  \label{Fig:Subnetwork}
\end{figure}

\subsubsection{Network Decomposition}\label{SEC:NetworkDecomposition}
To fully exploit the characteristics of our distance-based social
formation model, we modify the HC protocol so that the reduced
distance of an S--D pair can be carefully incorporated into the
protocol design. Owing to the central processor that leverages geo-information of nodes, it is possible to divide a network into multiple non-overlapping
subnetworks, each of which operates using the existing HC protocol
in~\cite{OzgurLeveque:TIT10} with a reduced network size
accordingly. When we denote $d_{s,v}$ as the distance between a
source $s$ and its destination $v$, we choose the side length
$l(n)$ of a subnetwork as $\mathbb{E}[d_{s,v}]n^{\epsilon}$ for an
arbitrarily small $\epsilon>0$ so that $s$ and $v$ can coexist
inside one subnetwork w.h.p., since the distance between an S--D
pair does not scale at a faster rate than $\mathbb{E}[d_{s,v}]$
within a factor of $n^{\epsilon}$ w.h.p. (refer to
Lemma~\ref{Lem:AverageDistanceMarkov}). The subnetworks of
constant side length
\begin{align}
l(n)=\mathbb{E}[d_{s,v}]n^{\epsilon} \label{EQ:side_length}
\end{align}
having S--D pairs are illustrated in Fig.~\ref{Fig:Subnetwork}.
There are $m:=nl^2(n)$ nodes on average in each subnetwork, where
the HC protocol in~\cite{OzgurLeveque:TIT10} is employed within
each subnetwork.

\subsubsection{Edge Node Problem and Protocol Refinement}\label{SEC:EdgeNodeProblem}
\begin{figure}[t!]
  \centering
  {
  \subfigure[Shifted to the right by $l(n)/2$]{
  \includegraphics[width=0.3\textwidth]{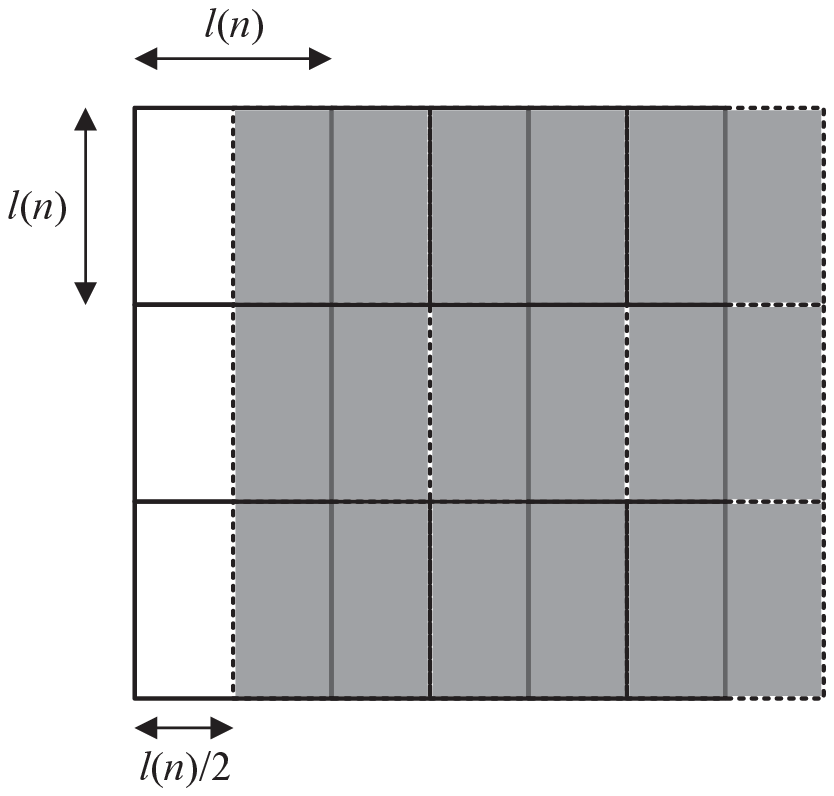}
  \label{FIG:Subnetwork-Shifted-Right}
  }
  \subfigure[Shifted up by $l(n)/2$]{
  \includegraphics[width=0.3\textwidth]{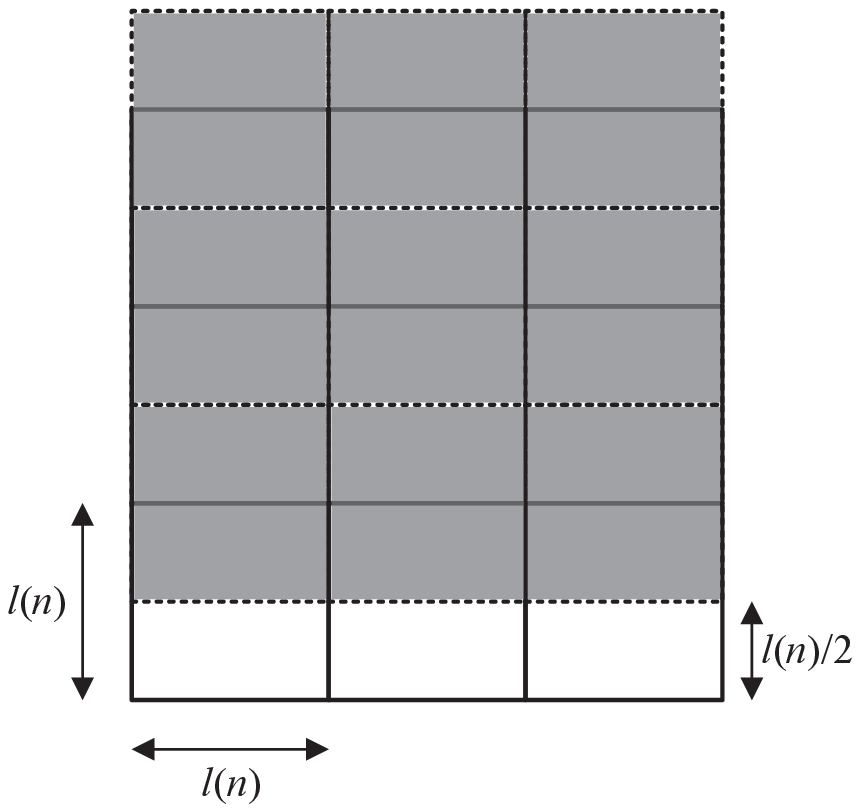}
  \label{FIG:Subnetwork-Shifted-Up}
  }
  \subfigure[Shifted diagonally by $\sqrt{2}l(n)/2$]{
  \includegraphics[width=0.3\textwidth]{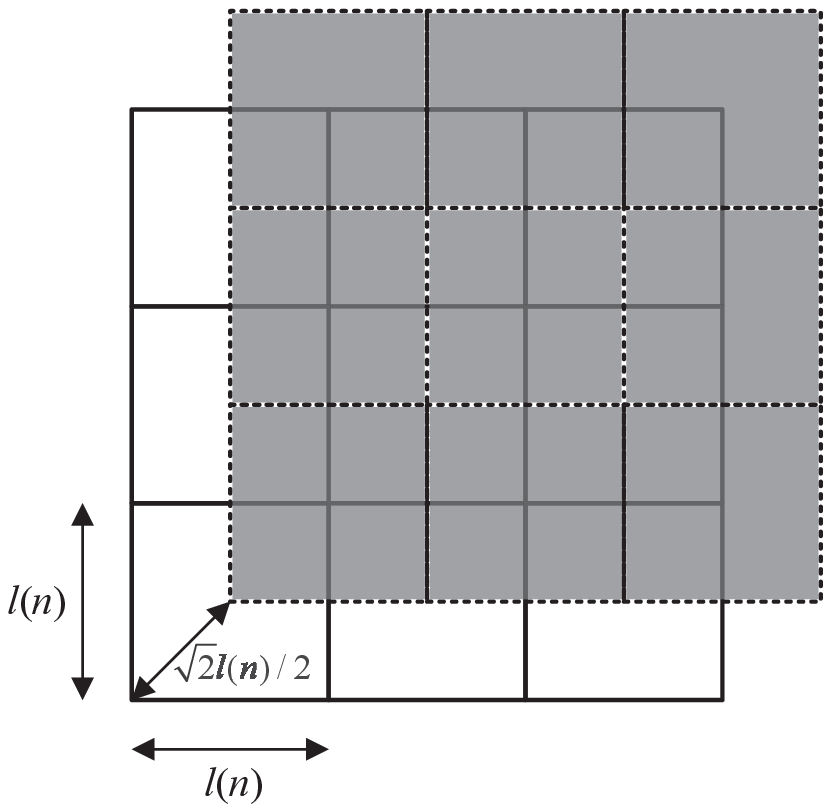}
  \label{FIG:Subnetwork-Shifted-Diagonal}
  }
  }
  \caption{The 4-TDMA operation of our network-decomposed HC protocol.}
  \label{FIG:Subnetwork-Shifted}
\end{figure}

When the whole network is decomposed into multiple subnetworks as
in Section~\ref{SEC:NetworkDecomposition}, there may exist some
sources in a subnetwork, whose destinations are outside of the
subnetwork, as depicted by red circles in
Fig.~\ref{Fig:Subnetwork}. Packets of these sources in the
subnetwork cannot be delivered to the corresponding destinations
using the existing HC framework that should be employed only
inside each subnetwork. In order to solve this edge node problem,
we apply a 4-TDMA operation with which all the subnetworks are
shifted to the right, shifted up by $l(n)/2$, and shifted
diagonally by $\sqrt{2}l(n)/2$ in consecutive time slots (see
Fig.~\ref{FIG:Subnetwork-Shifted}). By doing so, the packets of
the vast majority of S--D pairs can be successfully delivered in
one of time slots. Note that this subnetwork shift based on the
4-TDMA strategy does not fundamentally change the throughput and
delay scaling. In each time slot, the subnetworks operate in
parallel using the HC protocol with $\Theta(m)$ nodes in each
subnetwork.

\subsubsection{Network-Decomposed HC Protocol}
We assume that packets are conveyed by the network-decomposed HC
protocol {\em only} when the geographic distance between a source
node and its destination node is shorter than $\Theta(l(n))$
without affecting the overall throughput--delay trade-off. By
recalling Lemma~\ref{Lem:AverageDistanceMarkov}, we note that the
average number of S--D pairs whose distance is less than
$\Theta(l(n))$ is given by $\Theta(n)$ (refer to
Remark~\ref{Remark:LongDistance} for more details). According to
Sections~\ref{SEC:NetworkDecomposition}
and~\ref{SEC:EdgeNodeProblem}, the proposed network-decomposed HC
protocol is described as follows:
\begin{itemize}
  \item Step 1) Our ad hoc network is divided into multiple non-overlapping subnetworks, each of which has size $l(n)\times l(n)$.
  \item Step 2) The HC protocol in~\cite{OzgurLeveque:TIT10} is employed in each subnetwork.
  \item Step 3) All the subnetworks are shifted to the right by $l(n)/2$ compared to their original positions, and the HC protocol is employed in each shifted subnetwork.
  \item Step 4) All the subnetworks are shifted up by $l(n)/2$ compared to the original positions, and the HC protocol is employed in each shifted subnetwork.
  \item Step 5) All the subnetworks are shifted diagonally by $\sqrt{2}l(n)/2$ compared to the original positions, and the HC protocol is employed in each shifted subnetwork.
  \item Steps 2), 3), 4), and 5) are repeated in a TDMA manner.
\end{itemize}
Under the network-decomposed HC protocol, the throughput--delay
trade-off of a wireless {\em dense} network adopting the social
formation model is given by the expression in
(\ref{TD-Tradeoff-HC-Dense}) (refer to
Theorem~\ref{THM:TD-Tradeoff-HC}). It is shown that the
throughput--delay trade-off for the network-decomposed HC protocol
is improved as the social group density $\gamma$ increases, since
the number of subnetworks that are activated in parallel becomes
large with increasing $\gamma$.

\begin{remark}
In the original HC protocol, the average per-node transmit power
required to run the HC protocol in dense networks is $\frac{P}{n}$
but not $P$. This is because when the $\Theta(n)\times\Theta(n)$
long-range MIMO transmission is performed at the top level of the
hierarchy, the average distance between two clusters (i.e., the
transmitting nodes and the receiving nodes) is $O(1)$ and an array
gain of $n$ can be obtained. On the other hand, when the
network-decomposed HC protocol is employed in our dense network,
the average per-node transmit power becomes
\begin{align}\label{EQ:Avg-Tx-Power-HC-Dense}
  \frac{P}{m}l(n)^{\alpha}=\frac{P}{n}l(n)^{\alpha-2}
\end{align}
due to the fact that the network-decomposed HC protocol operates
within a subnetwork of $m$ nodes and the average distance between
two clusters is given by $O(l(n))$.
\end{remark}

\begin{remark}\label{Remark:LongDistance}
Our network-decomposed HC protocol operates within each subnetwork
in parallel. However, from Lemma~\ref{Lem:AverageDistanceMarkov},
there exists a non-zero probability that a relatively small
fraction of S--D pairs are further apart than the size of the
subnetwork. In our work, we assume that the packets of such S--D
pairs are conveyed by the MH protocol. It is not difficult to show
that the resulting throughput--delay trade-off does not change at
all by performing MH for such pairs.
\end{remark}

\subsection{Extended Network Configuration}\label{SEC:ExtendedNetwork}
In the previous subsection, we have focused on the analysis for a
dense network of unit area. We now turn to analyzing an extended
network of unit node density, whose size is
$\sqrt{n}\times\sqrt{n}$. Since the distance between a source and
its destination is increased by a factor of $\sqrt{n}$ in the
extended network, the received signal power will be decreased by a
factor of $n^{\alpha/2}$. Hence, the extended network of size
$\sqrt{n}\times\sqrt{n}$ can be treated as a dense network of size
$1\times 1$ with the reduced average per-node transmit power
constraint of $\frac{P}{n^{\alpha/2}}$. Thus, if the average
per-node transmit power required to perform the network-decomposed
HC protocol in the dense network is bounded by
$\frac{P}{n^{\alpha/2}}$, then the same throughput--delay
trade-off for the network-decomposed HC protocol can be achieved
as in the dense network case. Since the average per-node transmit
power depends on side length $l(n)$
in~\eqref{EQ:Avg-Tx-Power-HC-Dense}, which varies according to the
operating regimes, we compare the average per-node transmit power
with $\frac{P}{n^{\alpha/2}}$ in each regime as follows. Note that
this network transformation along with scaling parameters $\gamma$
and $\alpha$ is not a straightforward extension
of~\cite{OzgurLevequeTse:07}.

In Regime A (i.e., the low social group density regime), using
Lemma~\ref{Lem:AverageDistance}, (\ref{EQ:side_length}), and
(\ref{EQ:Avg-Tx-Power-HC-Dense}), the average per-node transmit
power for the network-decomposed HC protocol becomes
$\frac{P}{n^{1-\epsilon}}$ in our dense network. In order to
satisfy the equivalent power constraint $\frac{P}{n^{\alpha/2}}$,
we need a bursty transmission strategy similarly as
in~\cite{OzgurLevequeTse:07}. Specifically, the network-decomposed
HC protocol is performed during a fraction
$\frac{1}{n^{\alpha/2-1+\epsilon}}$ of the time with per-node
power $\frac{P}{n^{1-\epsilon}}$ and remains silent for the rest
of the time. Then, the throughput is reduced to
$T(n)=\Theta\left(n^{b-\alpha/2+1-\epsilon}\right)$ while the
delay is the same as that in the dense network. In Regime B (i.e.,
the medium social group density regime), the average per-node
transmit power required to run the network-decomposed HC protocol
becomes $\frac{P}{n^{1+(\gamma/2-1)(\alpha-2)-\epsilon}}$ in the
dense network due to the fact that $l(n)$ depends on the social
group density $\gamma$. In the regime, we need to use a bursty
transmission that runs the network-decomposed HC protocol during a
fraction $\frac{1}{n^{(1-\alpha/2)(\gamma-3)+\epsilon}}$ of the
time, resulting in
$T(n)=\Theta\left(n^{b-(\gamma-2)(b-1)-(1-\alpha/2)(\gamma-3)-\epsilon}\right)$
with the same delay as in the dense network. In Regime C (i.e.,
the high social group density regime), the average per-node
transmit power is $\frac{P}{n^{\alpha/2-\epsilon}}$ in the dense
network due to $l(n)=\Theta(n^{-1/2+\epsilon})$. Hence, the
network-decomposed HC protocol runs during a fraction
$n^{-\epsilon}$. The throughput is then given by
$T(n)=\Theta(n^{1-\epsilon})$ with the same delay as in the dense
network.

Based on this bursty modification according to each operating
regime, we can establish the throughput--delay trade-off in
(\ref{EQ:TD-Tradeoff-HC-Extended}) for the network-decomposed HC
protocol in the extended network with our social formation model
(see Theorem~\ref{THM:TD-Tradeoff-HC-Extended}). Compared to the
dense network case, the throughput--delay trade-offs in Regimes A
and B are degraded as the path-loss exponent $\alpha$ increases
since the extended network is power-limited in these two regimes.
On the other hand, in Regime C, the network is not power-limited
due to the high social group density, where the same
throughput--delay trade-off can be achieved as in the dense
network case.

\section{Numerical Evaluation}\label{SEC:Simulation}
In this section, we perform computer simulations
according to finite values of $n$ to validate that our analytical
results show trends consistent with numerical results. We evaluate
the performance of the network-decomposed HC protocol in terms of
the total throughput $T(n)$, which can be computed as the sum of
the transmission rates in parallel across all subnetworks. The
capacity of each subnetwork is bounded by the rate of long-range
MIMO transmission between two clusters having a source and its
destination in a subnetwork. A sufficient number of nodes
($n=256$) are deployed so that a large-scale network is suitably
modeled in practice. It is assumed that the network size is given
by $100 \times 100$ (m$^2$). All the transmit power is set to the
same value as the noise variance.

Now, we turn to describing how to generate
multiple network configurations, each of which has different
numbers of subnetworks, depending on values of the social group
density $\gamma$. Let $b_\gamma$ denote the parameter $b$ in
Theorems 2 and 4 for given $\gamma$. In our simulations, we set
$b_2=\frac{1}{4}$. Then, from the results of the two theorems, we
have $b_{\gamma}=\frac{b_2}{3-\gamma}=\frac{1}{4(3-\gamma)}$ to
ensure that the delays for all values of $\gamma$ are the same in
order sense. From the fact that the side length of a subnetwork in
(12) is given by $\mathbb{E}[d_{s,v}]$ in (3) for given $\gamma$,
it follows that four values $2$, $2.25$, $2+\frac{\log_2 3}{4}$,
and $2.5$ for $\gamma$ are chosen to generate such network
configurations that have 1, 4, 9, and 16 subnetworks,
respectively. Note that $\gamma=2$ corresponds to the baseline
employing the conventional HC protocol since there exists a single
subnetwork.

\begin{figure}[t!]
  \centering
  \includegraphics[width=0.6\textwidth]{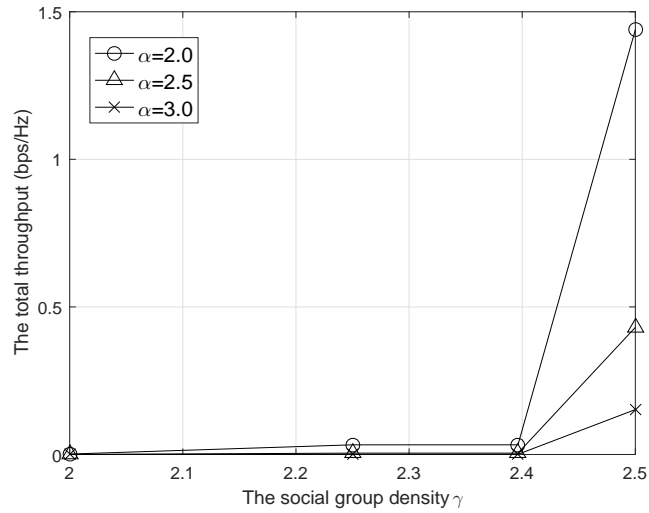}\\
  \caption{The total throughput according to the social group density $\gamma$, where $\alpha\in\{2.0, 2.5, 3.0\}$.}
  \label{Fig:Simulation}
\end{figure}

In Fig.~\ref{Fig:Simulation}, the total
throughput versus the social group density $\gamma$ is
illustrated, where $\alpha\in\{2, 2.5, 3\}$. From the figure, the
following observations are made under our simulation environments:
the throughput performance is degraded with increasing path-loss
exponent $\alpha$ due to more severe path-loss attenuation; the
total throughput is gradually enhanced as $\gamma$ increases up to
2.4; and the total throughput is greatly improved when
$\gamma=2.5$ since the number of subnetworks is largely increased
as the distance between a source and its destination,
$\mathbb{E}[d_{s,v}]$, decreases with respect to $\gamma$. This
result implies that the social behavior among nodes in ad hoc
networks has a significant impact on the throughput performance.

\section{Concluding Remarks}\label{SEC:Conclusion}
By introducing a new HC protocol, we completely characterized the
general throughput--delay trade-off of dense and extended ad hoc
networks with the distance-based social formation model
parameterized by the social group density $\gamma$ and the number
of social contacts per node, $q$, where a source selects one of
its social contacts as its destination uniformly at random. More
precisely, we proposed the network-decomposed HC protocol so that
the networks operate properly under our social formation model in
terms of maximizing the throughput--delay trade-off. To more
concisely show our main results, we also identified three
operating regimes on the throughput--delay trade-off with respect
to $\gamma$ and $q$. In the dense network, we showed that when
$\gamma$ is small, the throughput--delay trade-off is the same as
the non-social behavior scenario; on the other hand, when $\gamma$
increases, the throughput--delay trade-off is significantly
improved; and when $\gamma$ becomes large, the maximum throughput
$\Theta(n)$ can be achieved via a single-hop transmission, leading
to the delay of $\Theta(1)$. In addition, we analyzed the
corresponding throughput--delay trade-off in the extended network
through the nontrivial network transformation strategy, and
investigated the operating regimes such that the
network-decomposed HC protocol outperforms the MH protocol
according to parameters $\gamma$ and $\alpha$. Suggestions for
further research include analyzing performance on the throughput
and delay when the size of social groups follows a well-known
Zipf's distribution instead of assuming the same size of all
social groups. Another interesting direction is
to investigate the effect of queuing delay on the
throughput--delay trade-off for the network-decomposed HC protocol
in ad hoc networks with social relationships.

\appendices
\renewcommand\theequation{\Alph{section}.\arabic{equation}}
\setcounter{equation}{0}

\section{Proof of Theorem \ref{THM:TD-Tradeoff-HC}}\label{SEC:Proof-1}
In a dense network, the throughput--delay trade-off achieved by
the network-decomposed HC protocol in a subnetwork consisting of
$m=nl^2(n)$ nodes is given by~\cite{OzgurLeveque:TIT10}
\begin{align}
(T(n),D(n))=\Theta(m^{b}/\log m, m^b\log m), \nonumber
\end{align}
where $0\leq b<1$. Since there are $\frac{n}{m}$ subnetworks over
the whole network, the aggregate throughput is $\frac{n}{m}$ times
the throughput of each subnetwork while the delay of the network
remains the same due to the fact that all the subnetworks operate
in parallel. Hence, the throughput--delay trade-off of the network
is given by
\begin{align}\label{Eq:HC-Delay-Throughput-Social}
(T(n),D(n))=\Theta(m^{b-1}n/\log m, m^b\log m),
\end{align}
where $0\leq b<1$. Substituting
$m=n\mathbb{E}^2[d_{s,v}]n^{2\epsilon}$ into
(\ref{Eq:HC-Delay-Throughput-Social}), where $d_{s,v}$ denotes the
distance between a source $s$ and its destination $v$, we have
\begin{align}
    (T(n),D(n))&=\Theta(n^b(\mathbb{E}[d_{s,v}]n^{\epsilon})^{2b-2}/\log (n(\mathbb{E}[d_{s,v}]n^{\epsilon})^2),
    \nonumber\\
    &~~~~~~~
    n^b(\mathbb{E}[d_{s,v}]n^{\epsilon})^{2b}\log
    (n(\mathbb{E}[d_{s,v}]n^{\epsilon})^2)).\nonumber
\end{align}
In what follows, we derive the throughput and the delay according
to each operating regime.

\subsection{Regime~A}
For $q(n)=\omega(1)$, since $\mathbb{E}[d_{s,v}]=\Theta(1)$ from
Lemma~\ref{Lem:AverageDistance}, the throughput $T(n)$ and the
delay $D(n)$ are given by
\begin{align}
(T(n),D(n))&=\Theta(n^{b+\epsilon(2b-2)}/\log n^{1+2\epsilon}, n^{b+2\epsilon b}\log n^{1+2\epsilon})
\nonumber\\
   &=\Theta(n^{b-\epsilon}, n^{b+\epsilon}).\nonumber
\end{align}
For $q=\Theta(1)$ and $0\leq\gamma<2$, we have the same result as
above due to $\mathbb{E}[d_{s,v}]=\Theta(1)$.

\subsection{Regime~B}
For $q=\Theta(1)$ and $2\leq\gamma\leq 3$, it follows that
$\mathbb{E}[d_{s,v}]=\Theta\left(\left(\frac{\log
n}{n}\right)^{\frac{\gamma}{2}-1}\right)$ from
Lemma~\ref{Lem:AverageDistance}. Thus, we have
\begin{align} \nonumber
    T(n)    
    =\Theta\left(n^b\left(\left(\frac{\log n}{n}\right)^{\frac{\gamma}{2}-1}n^{\epsilon}\right)^{2b-2}
    \frac{1}{\log \left(n\left(\left(\frac{\log n}{n}\right)^{\frac{\gamma}{2}-1}n^{\epsilon}\right)^{2}\right)}\right),
    \nonumber
\end{align}
\begin{align}
    D(n)
    =\Theta\left(n^b\left(\left(\frac{\log n}{n}\right)^{\frac{\gamma}{2}-1}n^{\epsilon}\right)^{2b}
    \log\left(n\left(\left(\frac{\log
    n}{n}\right)^{\frac{\gamma}{2}-1}n^{\epsilon}\right)^{2}\right)
    \right),
    \nonumber
\end{align}
which can be simplified to
$(T(n),D(n))=(\Theta\left(n^{b-(\gamma-2)(b-1)-\epsilon}\right),\Theta\left(n^{(3-\gamma)b+\epsilon}\right))$.

\subsection{Regime~C}
For $q=\Theta(1)$ and $\gamma>3$, it follows that
$\mathbb{E}[d_{s,v}]=\Theta\left(\sqrt{\frac{\log n}{n}}\right)$
from Lemma~\ref{Lem:AverageDistance}. Thus, we have
\begin{align}
    T(n)&=\Theta\left(n^b\left(\sqrt{\frac{\log n}{n}}n^{\epsilon}\right)^{2b-2}\frac{1}{\log \left(n\left(\sqrt{\frac{\log n}{n}}n^{\epsilon}\right)^{2}\right)}\right)
    \nonumber\\
    D(n)&=\Theta\left(n^b\left(\sqrt{\frac{\log n}{n}}n^{\epsilon}\right)^{2b}\log\left(n\left(\sqrt{\frac{\log
    n}{n}}n^{\epsilon}\right)^{2}\right)\right), \nonumber
\end{align}
which can be rewritten as
$(T(n),D(n))=(\Theta\left(n^{1-\epsilon}\right),\Theta\left(n^{\epsilon}\right)$.
In consequence, the throughput--delay trade-off for the
network-decomposed HC protocol in the dense network is given by
the expression in (\ref{TD-Tradeoff-HC-Dense}), which completes
the proof of this theorem.

\section{Proof of Theorem \ref{THM:TD-Tradeoff-HC-Extended}}\label{SEC:Proof-2}
In an extended network, we derive the throughput--delay trade-off
achieved by the network-decomposed HC protocol according to each
operating regime in the following. Based on the power limitation
argument in Section~\ref{SEC:ExtendedNetwork}, the extended
network is equivalent to the dense network with the average
per-node transmit power constraint of $\frac{P}{n^{\alpha/2}}$.
Note that from (\ref{EQ:Avg-Tx-Power-HC-Dense}), the average
per-node transmit power required for the network-decomposed HC
protocol in the dense network is
$\frac{P}{m}l(n)^\alpha=\frac{P}{n}l(n)^{\alpha-2}$ (but not $P$).
We will compare this required power with the average per-node
power constraint of $\frac{P}{n^{\alpha/2}}$ in each regime.

\subsection{Regime~A}
For $q(n)=\omega(1)$ or
$\{q=\Theta(1)~\text{and}~0\leq\gamma<2\}$, the network-decomposed
HC protocol is performed with the average per-node transmit power
constraint of
$\frac{P}{n}l(n)^{\alpha-2}=\frac{P}{n}(n^{\epsilon})^{\alpha-2}$.
We use a bursty transmission strategy similarly as
in~\cite{OzgurLevequeTse:07} in order to satisfy the corresponding
power constraint $\frac{P}{n^{\alpha/2}}$, where the
network-decomposed HC protocol is performed during a fraction
$\frac{1}{n^{\alpha/2-1+\epsilon}}$ of the time with per-node power $\frac{P}{n^{1-\epsilon}}$ and remains
silent for the rest of the time. The throughput $T(n)$ and the
delay $D(n)$ are then given by
$T(n)=\Theta\left(n^{b-\alpha/2+1-\epsilon}\right)$ and
$D(n)=\Theta\left(n^{b+\epsilon}\right)$, respectively.

\subsection{Regime~B}
For $q=\Theta(1)$ and $2\leq\gamma\leq 3$, we use a bursty
transmission that runs the network-decomposed HC protocol during a
fraction $\frac{1}{n^{(1-\alpha/2)(\gamma-3)+\epsilon}}$ of the
time. In this case, we have
\begin{align}
   (T(n),D(n))=\left(\Theta\left(n^{(b-\alpha/2)(3-\gamma)+1-\epsilon}\right),\Theta\left(n^{(3-\gamma)b+\epsilon}\right)\right).
   \nonumber
\end{align}

\subsection{Regime~C}
For $q=\Theta(1)$ and $\gamma>3$, we use a bursty transmission
that runs the network-decomposed HC protocol during a fraction
$n^{-\epsilon}$, resulting in
$(T(n),D(n))=\left(\Theta\left(n^{1-\epsilon}\right),\Theta\left(n^\epsilon\right)\right)$.

In consequence, the throughput--delay trade-off for the
network-decomposed HC protocol in the extended network is given by
the expression in (\ref{EQ:TD-Tradeoff-HC-Extended}), which
completes the proof of this theorem.


\end{document}